\documentclass[preprint,showpacs,PRE,eqsecnum,amsmath,amssymb]{revtex4}

\usepackage{graphicx}

\global\arraycolsep=2pt         % Fix space around '=' in eqnarray's

\begin{document}

\title{Highly Charged Ions in a Dilute Plasma: \\ 
         An Exact Asymptotic Solution Involving Strong Coupling }

\author{Lowell S. Brown} \author{David C. Dooling} \author{Dean L. Preston}

\affiliation{
Los Alamos National Laboratory
\\
Los Alamos, New Mexico 87545
\\}

\date{\today}

\begin{abstract}

The ion sphere model introduced long ago by Salpeter is placed in
a rigorous theoretical setting.  The leading corrections to this
model for very highly charged but dilute ions in thermal
equilibrium with a weakly coupled, one-component background
plasma are explicitly computed, and the subleading corrections
shown to be negligibly small.  This is done using the effective
field theory methods advocated by Brown and Yaffe. Thus,
corrections to nuclear reaction rates that such highly charged
ions may undergo can be computed precisely.  Moreover, their
contribution to the equation of state can also be computed with
precision.  Such analytic results for very strong coupling are
rarely available, and they can serve as benchmarks for testing
computer models in this limit.

\end{abstract}

\pacs{{\bf 05.20.-\bf y}, 11.10.Wx, {\bf 52.25.-\bf b}}

\maketitle

\newpage

\section{Introduction and Summary}

Here we describe a plasma configuration whose exact asymptotic
solution can be obtained in a strong coupling limit.  The
solution is given by the ion sphere result presented by Salpeter
\cite{Sal} plus a simple smaller correction.  This is
accomplished by using the effective plasma field theory methods
advocated by Brown and Yaffe \cite{BY}.  In this field-theory
language, the old Salpeter result corresponds to the tree
approximation and our new correction is the one-loop term.  In
usual perturbative expansions, the tree approximation provides
the first, lowest-order term for weak coupling. Here, on the
contrary, the tree approximation provides the leading term for
strong coupling, with the corrections of higher order in the {\it
inverse coupling}. This is the only example of which we are aware
in which the tree approximation yields the strong coupling limit.
This strongly coupled system is interesting from a theoretical
point of view and our results can be used to check numerical
methods.

The plasma consists of very dilute ``impurity'' ions of very high
charge $Z_p e$, $Z_p \gg 1$, in thermal equilibrium with a classical,
one-component ``background'' plasma of charge $ze$ and number density
$n$, at temperature $T = 1/\beta$.  The background plasma is
neutralized in the usual way, and it is dilute.  We use rationalized
electrostatic units and measure temperature in energy units so that
the background plasma Debye wave number appears as 
\begin{equation}
 \kappa^2 = \beta \, (ze)^2 \, n \,.
\end{equation}
The internal  coupling of the background plasma is described by
the dimensionless coupling parameter
\begin{equation}
 g = \beta {(ze)^2 \over 4\pi} \, \kappa 
   = { (ze)^2 \over 4\pi T } \, \kappa \,.
\label{gee}
\end{equation}
The assumed weak coupling of the dilute background plasma is
conveyed by
\begin{equation}
g \ll 1 \,.
\end{equation}
Although the internal coupling of the background plasma to itself
is assumed to be very weak and the impurity ions are assumed to
be so very dilute that their internal interactions are also very 
small, we shall require that the ionic charge $Z_p$ is so great 
that the coupling between the impurity ions and the background 
plasma is very large.  To make this condition more precise, we
define 
\begin{equation}
\bar Z_p = { Z_p \over z } \,,
\end{equation}
which is the magnitude of the impurity charge measured in units
of the dilute background ionic charge. Then the explicit
condition that we require is that
\begin{equation} 
 g \bar Z_p \gg 1 \,.
\end{equation}

Since the limit that we use may appear to be obscure, we pause to
clarify it.  Even  though $g \bar Z_p \gg 1$, we assume that $g$
is sufficiently small that $g^2 \bar Z_p \ll 1$.  We may, for
example, take $g \to 0$ with $g^\alpha \, \bar Z_p = {\rm const.} $,
and $\alpha$ in the interval $ 1 < \alpha < 2$.  Then
$g \bar Z_p = {\rm const.} / g^{\alpha -1} \gg 1 $ while 
$g^2 \bar Z_p =  {\rm const.} \, g^{2-\alpha} \ll 1$.

Standard methods express the grand canonical partition function
in terms of functional integrals.  Brown and Yaffe \cite{BY} do
this, introduce an auxiliary electrostatic potential, and
integrate out the charged particle degrees of freedom to obtain
the effective theory. This technique will be described in more
detail in Sec.~\ref{rem} below. The saddle point expansion of
this form for the grand partition function yields a perturbative
expansion, with the tree approximation providing the lowest-order
term.  Here, on the contrary, we express the {\it impurity ion
number} in terms of an effective field theory realized by a
functional integral.  The saddle point of this form of the
functional integral involves a classical field solution driven by
a strong point charge.

The result for the impurity ion number reads
\begin{eqnarray}
&&N_p =  N_p^{(0)} \, \exp\Bigg\{ {3 \over 10 } \, (3g)^{2/3} \,
 \bar Z_p^{5/3}
 \nonumber\\
&+&  \!\!
\left({\,9\,\over g}\right)^{1/3} \!\! {\cal C} \, \bar Z_p^{2/3}
+ \cdots
- {1\over3} \, g \, \bar Z_p + {\cal O}(g^2 \bar Z_p) \Bigg\} .
\label{num}
\end{eqnarray}
Here $N_p^{(0)} \sim \exp\{ \beta \mu_p \} $ is the number of
impurity ions defined by the chemical potential $\mu_p$ in the
absence of the background plasma; keeping this chemical potential
fixed, the background plasma alters this number to be $N_p$.  The
added $\cdots$ stand for corrections to the analytical
evaluation of the classical action displayed in the 
$\bar Z_p^{5/3}$ and $\bar Z_p^{2/3}$ terms of Eq.~(\ref{num}).  
The sizes of these omitted corrections are compared to
the exact numerical evaluation of the action in
Fig.~\ref{ratio} below.  This figure shows that the relative
sizes of these terms are small ($ \ll 1$) in the limit in which we
work ($ gZ \gg 1 $). The constant ${\cal C} =
0.8499 \cdots$.  The final $ - g \bar Z_p / 3$ term in the exponent is
the relatively small one-loop correction.  As shown in detail in
the discussion leading to Eq.~(\ref{correct}) below, the error in
the result (\ref{num}) is of the indicated order $g^2 \, \bar Z_p = g
\, (g\bar Z_p)$ and is thus negligible in the limit $g \ll 1$ that
concerns us.
 
The number correction (\ref{num}) can be used to construct the grand
canonical partition function ${\cal Z}$ for the combined system by
integrating the generic relation 
\begin{equation}
 N_{a} = {\partial \over \partial \, \beta \mu_{a}} \, \ln {\cal Z} \,
\label{number}
\end{equation}
for $a = p$ and using the boundary condition that $N_{p} \rightarrow 0$ as
$\beta \mu_{p} \rightarrow -\infty$.
Since $N_p$ depends upon the chemical potential $\mu_p$ only in
the factor $N_p^{(0)} \sim \exp\{\beta \mu_p\}$, this integration gives
\begin{equation}
\ln{\cal Z} = N_{p} + N^{(0)} \,.
\label{solution}
\end{equation}
Here we have identified the constant of integration, the constant
that remains when $N_p$ vanishes, to be $N^{(0)}$, the number of
background plasma particles in the absence of the impurity
ions. In our limit in which the background plasma is very weakly
coupled, $N^{(0)} \sim \exp\{\beta \mu \}$ is just the number of
non-interacting particles of chemical potential $\mu$.

The equation of state can be found from the well-known relation
for a grand canonical ensemble with partition function ${\cal Z}$, 
\begin{equation}
 \beta p V = \ln{\cal Z} \,.
\label{pressure}
\end{equation}
However, the grand canonical partition function ${\cal Z}$ is a
function of the temperature and chemical potentials and, to
obtain the equation of state, we must re-express it in terms of
the observed, physical particle numbers rather than their
chemical potentials. 

To do this, we need to express 
$N^{(0)} \sim \exp\{\beta \mu \}$
in terms of the true
number of background particles $N$, a number that differs from
$N^{(0)}$ because of the presence of the impurity ions. There is
a significant difference because, although the impurity ions are
few in number, they are assumed to be extremely highly charged. 
We again use the general formula (\ref{number}), but this time to
compute $N$ using the solution (\ref{solution}):
\begin{equation}
N = {\partial N_p \over \partial \, \beta\mu} + N^{(0)} \,.
\end{equation}
The measured impurity number $N_{p}$ does depends upon 
$\beta \mu$ because it entails the dimensionless coupling
parameter $g$ defined in Eq.~(\ref{gee}).  For simplicity of
exposition, in that definition we used a Debye wave number 
$\kappa$ that was defined in terms of the true background density
$n$. Although the distinction is not important for the leading
terms that concern us, we nevertheless note that the correct 
wave number that appears in our functional
integral formalism involves the `bare' number density 
$n^{(0)} = N^{(0)} / V $,
with $g \sim \sqrt{n^{(0)}} \sim \exp\{ \beta \mu /2 \}$, and so
\begin{equation}
{\partial g \over \partial \beta \mu } = {1 \over 2} \, g \,.
\end{equation}
Hence,
\begin{equation}
N = N^{(0)} + {1\over2} \, g {\partial \over \partial g} \, N_p
\,.
\end{equation}

Using this relation to determine $N^{(0)}$ in terms of the 
physical quantities $N$ and $N_{p}$ places relationship 
(\ref{pressure}) of the pressure to the partition function
(\ref{solution}) in the proper form of an equation of state.  
To simply bring out the main point,
we include here only the leading terms, to obtain
\begin{eqnarray} p V &\simeq&
\left\{ N - \bar Z_p { (3g\bar Z_p)^{2/3} \over 10} \,
                 \, N_p \right\} \, T \,.
\end{eqnarray}
Although the fraction of impurity ions in the plasma
$ N_p / N$ may be quite small, there may be a significant
pressure modification if $\bar Z_p$ is very large.
Note that the free particle contribution, an additional term
of $N_p$, is omitted here since it is not multiplied by the large
factor in the term that we have retained.

The number result (\ref{num}) also directly yields the plasma 
correction to a nuclear fusion rate, since  
\begin{equation}
\Gamma = \Gamma_C \, { N^{(0)}_1 \over N_1} \,
{ N^{(0)}_2 \over N_2 } \,
 { N_{1+2} \over N^{(0)}_{1+2} } \,,
\label{xrated}
\end{equation}
where $\Gamma_C$ is the nuclear reaction rate for a thermal,
Maxwell-Boltzmann distribution of the initial (1,2) particles in
the absence of the background plasma.  We use the notation $1+2$
to denote an effective particle that carries the charge $(Z_1 +
Z_2) e $.  This formula was obtained in a different guise by
DeWitt, Graboske, and Cooper \cite{DGC}. The relation of the form
(\ref{xrated}) that we use to previous results is discussed in
detail in the Appendix.  The formula holds when the Coulomb
barrier classical turning point of the nuclear reaction is small
in comparison with the plasma Debye length.  This is spelled out
in detail in a recent work by Brown, Dooling, and Preston
\cite{BDP} who also show that the result (\ref{xrated}) is valid
even if the background plasma involves quantum corrections.  The
conditions needed for the formula (\ref{xrated}) to hold are also
discussed in the work of Brown and Sawyer \cite{BS}, although
sometimes in a rather implicit fashion. This work does show,
however, that the result (\ref{xrated}) is valid if $ \kappa \,
r_{\rm max} \ll 1$, where $\kappa = \beta e^2 n $ is the Debye
wave number and $r_{\rm max}$ is the turning point radius defined
by $ r_{\rm max} = 2 ( e^2 /4\pi \, m \omega^2)^{1/3} $ where $
\omega = 2\pi \, T / \hbar$ is the imaginary time frequency
associated with the temperature $T$. It should be remarked that
DeWitt, Graboske, and Cooper \cite{DGC} assumed that the nuclear
reaction rate formula (\ref{xrated}) held only if the background
plasma had a classical character, but that the work of Brown,
Dooling, and Preston \cite{BDP} shows that it is valid even if
the plasma involves quantum effects.
 
Our result (\ref{num}) for the number corrections presents the
plasma correction to the fusion rate for our special case as
\begin{eqnarray}
&& \Gamma   =   \Gamma_C  \exp\left\{ {3\over10} \, (3 g)^{2/3} 
\left[ \left( \bar Z_1 \! + \! \bar Z_2 \right)^{5/3} 
\!\! - \bar Z_1^{5/3} \!\! - \bar Z_2^{5/3} \right] \right\}  
\nonumber\\
 && \,\,
 \exp\left\{ \left({\,9\, \over g}\right)^{1/3} {\cal C}
\left[ \left(\bar Z_{1}+\bar Z_{2}\right)^{2/3} - \bar Z_{1}^{2/3}
- \bar Z_{2}^{2/3} \right] \right\} \,.
\nonumber\\
&&.
\end{eqnarray}
The first line agrees with Salpeter's calculation \cite{Sal}; 
the second is new. Again the correction can be large.

We turn now to describe the basis for these results in detail.

\section{Remembrance of Things Past}
\label{rem}

To begin, we need to review a simple case of the general plasma
effective field theory formulation presented by 
Brown and Yaffe \cite{BY}.  First we note that the grand
canonical partition function for a one-component classical
plasma may be expressed as the functional
integral (which are discussed in detail, for example, in
  the first chapter of the book by Brown \cite{Brown}) ,
\begin{eqnarray}
{\cal Z} &=& \int [d\chi] \, \exp\Bigg\{ - \int (d^3{\bf r}) \Big[
{\beta \over 2} \, \Big( \nabla \chi({\bf r})  \Big)^2
\nonumber\\
&& \qquad\qquad\qquad
-g_S\,\lambda^{-3} e^{\beta\mu} \, e^{ize\beta \, \chi({\bf r})}
\Big] \Bigg\} \,.
\label{fun}
\end{eqnarray}
Here 
\begin{equation}
\lambda^{-3} = \int { (d^3{\bf p}) \over (2\pi\hbar)^3 }
\, \exp\left\{ - \beta \, {{\bf p}^2 \over 2m} \right\} 
\label{lambda}
\end{equation}
defines the thermal wave length $\lambda$ of the plasma particles
of mass $m$.  These particles have a chemical potential $\mu$ and
spin weight $g_S$ so that their density in the free-particle
limit is given by
\begin{equation}
n^{(0)} = g_S \lambda^{-3}\, e^{\beta\mu} \,.
\end{equation}
We use rationalized Gaussian units so that, for example, the 
Coulomb potential appears as $ \phi = e / 4\pi \, r$. 
We shall be a little cavalier about the uniform, rigid
neutralizing background that we tacitly assume to be present.  We
shall explicitly include its effects when needed.

The validity of the functional integral representation
(\ref{fun}) is easy to establish.  The second part in the
exponential is written out in a series so as to produce the
fugacity expansion
\begin{eqnarray}
{\cal Z} &=& \sum_{n=0}^\infty \, { 1 \over n!} \, 
\left( g_S \, \lambda^{-3} \right)^n \, e^{n \beta\mu} \,
\int (d^3{\bf r}_1) \cdots (d^3{\bf r}_n) \,
\nonumber\\
&& \qquad
\int [d\chi] \, \exp\Bigg\{ - \int (d^3{\bf r}) \big[
{\beta \over 2} \, \Big( \nabla \chi({\bf r})  \Big)^2
\nonumber\\
&& \qquad\qquad
+ i ze \beta \chi({\bf r}) \, \sum_{a=1}^n \,
   \delta ( {\bf r} - {\bf r}_a ) \Big] \Bigg\} \,.
\end{eqnarray}
This Gaussian functional integral can be performed by the
functional integration field variable translation
\begin{equation}
\chi({\bf r}) = \chi'({\bf r}) - \sum_{a=1}^n \,
  { i ze \over 4\pi \, | {\bf r} - {\bf r}_a | } \,.
\end{equation}
Since
\begin{equation}
- \nabla^2 \, {1 \over 4\pi \, | {\bf r} - {\bf r}_a | } 
= \delta( {\bf r} - {\bf r}_a ) \,,
\end{equation}
and the Laplacian $\nabla^2$ can be freely integrated by parts in
the quadratic form $ \chi (- \nabla^2 ) \chi$, after the
translation a Gaussian functional integration appears with
quadratic form $ \chi' ( - \nabla^2 ) \chi' $ with no coupling linear
in $\chi'$.  The original measure $[d\chi] = [d\chi']$ is taken
to include factors such that this remaining purely Gaussian
function integral is simply unity.  For pedagogical clarity, we
make use of the definition (\ref{lambda}) of the thermal
wavelength to write the result of these manipulations as
\begin{eqnarray}
{\cal Z} &=& \sum_{n=0}^\infty \, { 1 \over n!} \, 
g_S^n \, e^{n \beta\mu} \,
\int {(d^3{\bf r}_1) (d^3{\bf p}_1) \ \over (2\pi\hbar)^3 } \cdots 
{(d^3{\bf r}_n) (d^3{\bf p}_n) \over (2\pi\hbar)^3 } \,
\nonumber\\
&& 
 \exp\left\{ -\beta \left[ \sum_{a=1}^n \,
 { {\bf p}_a^2 \over 2m } + {1\over2} \, \sum_{a,b=1}^n \,
{(ze)^2 \over 4\pi | {\bf r}_a - {\bf r}_b | } \right]\right\} \,.
\nonumber\\
&&
\end{eqnarray}
This is precisely the familiar fugacity expansion of the
classical grand canonical partition function.  The diagonal sum
where $a=b$ in the Coulomb potential must be deleted.  This
omission of the infinite self-energy terms is automatic if the
dimensional regularization scheme is employed as advocated by
Brown and Yaffe \cite{BY}.  Here we shall instead regulate the theory
by (at first implicitly) replacing the point source 
$ \delta({\bf r} - {\bf r}_a)$ with a  source
$ \delta_{R}({\bf r} - {\bf r}_a)$ that has a small extent about 
${\bf r}_a$ and (at first implicitly) removing the self energy terms,
with the limit $\delta_R \to \delta$ finally taken in the subtracted theory. 

The derivative of the logarithm of a grand canonical partition
function with respect to a chemical potential (times $\beta$)
gives the particle number conjugate to that chemical
potential. Thus, if we temporarily add another particle species
$p$ of charge $e_p = Z_pe$ to the previous functional integral,
take the described derivative, and then take the limit in which
this new species is very dilute, we get the desired functional
integral representation for the background plasma correction to the
new species free particle number relation in the presence of plasma
interactions,  
\begin{eqnarray}
&& N_p = 
\nonumber\\
&&{ N_p^{(0)} \over {\cal Z} } 
\int [d\chi] e^{ i Z_p e \beta \chi({\bf 0})}
\exp\Bigg\{\! - \! \int (d^3{\bf r}) \Big[
{\beta \over 2} \, \Big( \nabla \chi({\bf r})  \Big)^2
\nonumber\\
&& \qquad
- n \, \Bigg( e^{ize\beta \, \chi({\bf r})} - 1 
- ize \beta \, \chi({\bf r}) \Bigg) \, 
\Big] \Bigg\} \,.
\label{clever}
\end{eqnarray}
To express this more precisely, in Eq.~(\ref{clever}) $N_p^{(0)}
= g_{S_p} \, \lambda_p^{-3} {\cal V} \, \exp\{ \beta\mu_p \} $,
where the subscript $p$ is used to indicate that these are the
properties of the sparsely populated `impurity' ions of charge
$e_p = Z_pe$, with ${\cal V}$ denoting the system volume.  So
Eq.~(\ref{clever}) describes the background plasma correction to
the free-particle chemical potential -- number relationship for
these $p$ ions immersed in the weakly-coupled, one-component
plasma.  The original chemical potential derivative that leads to
this result entailed a volume integral.  In virtue of the
translational invariance of the background plasma, the result is
independent of the particular value of the spatial coordinate in
the electric potential $\chi({\bf r})$ in the initial factor, and
this coordinate may be placed at the origin (as we have done),
giving the factor $ e^{ ie_p \beta \, \chi({\bf 0})} $ shown.
The volume integral then combines to form the total free-particle
number $N_p^{(0)}$ that appears as a prefactor.  We have now
subtracted terms from the second exponential, the exponential of
the action functional of the background plasma, to remove an
overall number contribution and to include the effect of the
rigid neutralizing background.  These same subtractions must now
be made in the normalizing partition function ${\cal Z}$ that
appears in the denominator of Eq.~(\ref{clever}).  Thus ${\cal
Z}$ is defined by the functional integral of the second 
exponential that appears in Eq.~(\ref{clever}). The effect of
the uniform neutralizing rigid background charge is contained in
the term $ize\beta\chi$ that is subtracted from the exponential
$\exp\{ize\beta\chi\}$. The additional $1$ is subtracted from this
exponential for convenience.

To simplify the notation, we write Eq.~(\ref{clever}) as simply
\begin{equation}
N_p = { N_p^{(0)} \over {\cal Z}} \, \int [d\chi] \, e^{-S[\chi]} \,,
\label{compact}
\end{equation}
where the effective action $S[\chi]$ contains all the terms in both
exponents in Eq.~(\ref{clever}).  The loop expansion is an expansion
about the saddle point  of the functional integral.  At this point,
the action $S[\chi]$ is stationary, and thus the field $\chi$ at this
point obeys the classical field equation implied by the stationarity
of the action. 

The tree approximation is given by the evaluation of 
$S[\chi]$ at the classical solution
\begin{equation}
\chi({\bf r})  \to i \phi_{\rm cl}({\bf r}) \,,
\end{equation}
namely
\begin{eqnarray}
&&S[i\phi_{cl}] = 
- \int (d^3{\bf r}) \Bigg\{
{\beta \over 2} \,\Big(\nabla \phi_{\rm cl}({\bf r}) \Big)^2
\nonumber\\
&&
+ n \left[ e^{- \beta ze \, \phi_{\rm cl}({\bf r})} - 1 
+  \beta ze \, \phi_{\rm cl}({\bf r}) \right] 
- \beta Z_pe \delta({\bf r}) \, \phi_{\rm cl}({\bf r}) 
 \Bigg\} \,,
\nonumber\\
&&
\label{action}
\end{eqnarray}
whose stationary point defines the classical field equation
\begin{equation}
- \nabla^2 \phi_{\rm cl} ({\bf r}) = z e n \left[ 
  e^{-\beta ze \phi_{\rm cl}({\bf r})} - 1 \right]
+ Z_p e \, \delta({\bf r}) \,.
\label{obey}
\end{equation}
This equation defining the classical potential 
$\phi_{\rm cl}({\bf r})$ is of the familiar Debye-H\"uckel form,
and it could have been written down using simple physical reasoning.
However, we have placed it in the context of a systematic
perturbative expansion in which the error of omitted terms can be
ascertained. In particular, we shall describe the one-loop
correction that is automatically produced by our formalism. 
Moreover, we shall prove that higher-order corrections may
be neglected. Our approach using controlled approximations
in which the error is assessed, and making precise evaluations 
of a well defined perturbative expansions in terms of correctly
identified coupling parameters, differs in spirit from much of 
the traditional work in plasma physics. For example, although
previous work has been done by Vieillefosse \cite{Vie} on the 
solution of the non-linear Debye-H\"uckel equation, this work was 
not done in the context of a systematic, controlled approximation.

The one-loop correction to this first tree approximation is
obtained by writing the functional integration variable as
\begin{equation}
\chi({\bf r}) = i \phi_{\rm cl}({\bf r}) + \chi'({\bf r}) \,,
\end{equation}
and expanding the total action in Eq.~(\ref{compact}) to quadratic
order in the fluctuating field $\chi'$.  Since $i \phi_{\rm cl}$
obeys the classical field equation, there are no linear terms in
$\chi'$ and we have, to quadratic order,
\begin{eqnarray}
 S[\chi] &=& 
 S[i\phi_{\rm cl}]
\nonumber\\\
&+& 
{ \beta \over 2} \, \int (d{\bf r}) \, 
  \chi'({\bf r}) \,  \left[ - \nabla^2 + 
\kappa^2 \, e^{ - \beta ze \, \phi_{\rm cl}({\bf r}) }
 \right] \,   \chi'({\bf r}) \,,
\nonumber\\
&&
\end{eqnarray}
where
\begin{equation}
\kappa^2 = \beta \, (ze)^2 \, n 
\end{equation} is the squared Debye wave number of the mobile
ions.  The resulting Gaussian functional integral produces an
infinite dimensional, Fredholm determinant.  In this same
one-loop order, the normalizing partition function ${\cal Z}$ 
is given by the same determinant except that it is evaluated at 
$\phi_{\rm cl} = 0$.  Hence, to tree plus one-loop order, 
\begin{equation}
N_p = N_p^{(0)} \, 
{ {\rm Det}^{1/2} \left[ - \nabla^2 + \kappa^2 \right] \over
{\rm  Det}^{1/2} \left[ - \nabla^2 + \kappa^2 \, 
    e^{-\beta ze \, \phi_{\rm cl}} \right] } \,
\exp\left\{- S[i \phi_{\rm cl}] \right\} \,.
\end{equation}

\section{Computation}

\subsection{Tree}

To solve the classical field equation (\ref{obey}) in the large 
$Z_p$ limit, we first note that
the classical potential must vanish asymptotically 
so as to ensure that the resulting total charge density vanishes
at large distances form the `external' point charge $e_p = Z_pe$,
\begin{equation}
|{\bf r}| \to \infty \,: \qquad
  e n \, \left[1 -  
e^{ - \beta ze \, \phi_{\rm cl}({\bf r}) }  \right] \to 0 \,.
\label{neut}
\end{equation}
Since $\phi_{\rm cl}$ vanishes asymptotically, its defining
differential equation (\ref{obey}) reduces at large distances to
\begin{equation}
- \nabla^2 \phi_{\rm cl}({\bf r}) \simeq 
- \kappa^2 \,  \phi_{\rm cl}({\bf r}) \,,
\end{equation}
and thus, for $|{\bf r}|$ large,
\begin{equation}
 \phi_{\rm cl}({\bf r}) \simeq ({\rm const}) \,
   { e^{- \kappa |{\bf r}| } \over |{\bf r}| } \,.
\label{large}
\end{equation}
Since this is exponentially damped, the coordinate integral of
the left-hand side of Eq.~(\ref{obey}) vanishes by Gauss'
theorem, and we obtain the integral constraint
\begin{equation}
 z \, n \, \int (d{\bf r}) \, 
\left[1 -  e^{-\beta ze \, \phi_{\rm cl}({\bf r}) } \right]
= Z_p \,.
\label{intc}
\end{equation}

For small $r \equiv |{\bf r}|$, the point source driving term in 
the classical field equation dominates, giving the Coulomb
potential 
solution
\begin{equation}
\phi_{\rm cl}({\bf r}) \simeq { Z_p e \over 4\pi \, r} \,.
\label{short}
\end{equation}
Thus we write
\begin{equation}
\phi_{\rm cl}({\bf r}) = { Z_p e \over 4\pi \, r} \, u(\xi) \,,
\label{udef}
\end{equation}
where
\begin{equation}
\xi = \kappa r \,,
\end{equation}
and the point driving charge $Z_pe$ is now conveyed in the boundary
condition 
\begin{equation}
u(0) = 1 \,.
\label{norm}
\end{equation}
The other boundary condition is the previously noted large $r$
limit (\ref{large}) which now appears as
\begin{equation}
\xi \to \infty \,: \qquad\qquad u(\xi) \sim e^{- \, \xi} \,.
\end{equation}

The action (\ref{action}) corresponding to the classical solution
is divergent since it includes the infinite self-energy of the
point charge $ e_p = Z_p e$ impurity.  This self-energy must be
subtracted to yield the finite, physical action.  Following
standard practice in quantum field theory, the divergent
classical action (\ref{action}) and the self-energy are first
regularized --- rendered finite --- by replacing the point charge
with a finite source.  The self-energy is then subtracted, and
finally the point source limit is taken.  Regularization is
achieved by the replacement 
$\delta({\bf r}) \to \delta_R({\bf r})$,
where $\delta_R({\bf r})$ is a smooth function of compact
support. The regularized action obtained by making this 
substitution in the action $S[i\phi_{\rm cl}]$ defined by
Eq.~(\ref{action}) will be denoted as $S_{\rm reg}$. The 
regularized self field 
$\phi_{{\rm self}}({\bf r})$ is the solution of
\begin{equation}
- \nabla^2 \, \phi_{{\rm self}}({\bf r}) = Z_pe \,
\delta_R({\bf r}) \,,
\end{equation}
and it defines the self-action 
\begin{eqnarray}
&&S_{\rm self} = 
\nonumber\\
&&
- \beta \int \left( d^3 {\bf r} \right) \, \left\{
{1\over2} \, \Big( \nabla \, \phi_{\rm self}({\bf r}) \Big)^2 
       - Z_pe \, \delta_R({\bf r}) \, \phi_{\rm self}({\bf r}) 
\right\} \,.
\nonumber\\
&&
\label{tocast}
\end{eqnarray}
The identity
\begin{equation}
\beta \int \left( d^3{\bf r} \right) \, \left\{ 
\Big( \nabla \phi_{{\rm self}}({\bf r}) \Big)^2 - Z_pe \,
 \delta_R({\bf r}) \, \phi_{{\rm self}}({\bf r}) \right\} = 0 \,,
\label{null}
\end{equation}
which is easily verified through partial integration and use of
the field equation obeyed by $\phi_{\rm self}$ can be used to
write the self-energy action (\ref{tocast}) as
\begin{eqnarray}
S_{\rm self} &=& 
 \beta \int \left( d^3 {\bf r} \right) \, 
{1\over2} \, \Big( \nabla \, \phi_{\rm self}({\bf r}) \Big)^2 
\nonumber\\
&=& \beta \int \left( d^3 {\bf r} \right) \, 
{1\over2} \, {\bf E}^2_{\rm self}({\bf r}) \,,
\label{alt}
\end{eqnarray}
which is just the impurity's field energy divided by the
temperature. 
It is convenient to use this form (\ref{alt}) in subtracting off the
self-energy from $S_{\rm reg}$ and to also subtract the identity  
\begin{equation}
\beta \int \left( d^3{\bf r} \right) \, \left\{ 
 \nabla \phi_{{\rm self}}({\bf r}) \cdot  \nabla \phi_{{\rm cl}}({\bf r})
- Z_pe \, \delta_R({\bf r}) \, \phi_{{\rm cl}}({\bf r}) \right\} = 0 \,,
\label{nulll}
\end{equation}
proved in the same manner as Eq.~(\ref{alt}). 
The point source limit
$ \delta_R({\bf r}) \to \delta({\bf r})$ can now be taken to secure
the well-defined result
\begin{eqnarray}
&& S[i\phi_{\rm cl}] \to S_{\rm sub}[i\phi_{\rm cl}] =
\nonumber\\&&
   - \beta \int \left( d^3{\bf r} \right) \, {1\over2} \, \left[
 \nabla \Big( \phi_{\rm cl}({\bf r}) - \phi^P_{\rm self}({\bf r})
      \Big) \right]^2 
\nonumber\\
&&  - n \int \left( d^3{\bf r} \right) \,
\left[ e^{- \beta ze \, \phi_{\rm cl}({\bf r})} - 1 
+  \beta ze \, \phi_{\rm cl}({\bf r}) \right] ,
\label{sub}
\end{eqnarray}
where
\begin{equation}
\phi^P_{\rm self}({\bf r}) = {Z_pe \over 4\pi \, r} 
\end{equation}
is the point-source limit of the self-field.

Using the form (\ref{udef}) for the classical solution we 
have, remembering that $u(0) = 1$, 
\begin{eqnarray}
&& 4\pi \, r^2 \,  \left[ \nabla 
   \Big( \phi_{\rm cl}({\bf r}) - \phi^P_{\rm self}({\bf r}) \Big) 
      \right]^2 
\nonumber\\
&& = {(Z_p e)^2 \over 4\pi} \, \left[ { du(r) \over dr}
-  {1 \over r} \, \Big( u(r) - u(0) \Big)  \right]^2
\nonumber\\
&& = {(Z_pe)^2 \over 4\pi} \, \left\{ \left( { du(r) \over dr} \right)^2 
- {d \over dr}  \, \left[ {1 \over r} \, \Big( u(r) - u(0) \Big)^2
   \right]  \right\} \,.
\nonumber\\
&&
\end{eqnarray}
The final total derivative that appears here gives a null result
since the end-point contributions vanish.  Hence
the subtracted action (\ref{sub}) now appears as
\begin{eqnarray}
&&S_{\rm sub}[i\phi_{\rm cl}] =
 - \int_0^\infty dr \Bigg\{ {\beta \over 2} \,
{ Z_p^2 e^2 \over 4\pi} \, \left( { du \over dr} \right)^2 
+ 4\pi r^2 \, n 
\nonumber\\
&& \qquad
\left[ 
\exp\left\{ - { \beta Z_p z e^2 \over 4\pi r} \, u \right\} - 1 
+ { \beta Z_p z e^2 \over 4\pi r} \, u \, \right] \Bigg\} \,.
\end{eqnarray}
Changing variables to $\xi = \kappa r$ and using the previously
defined plasma coupling constant
$
g =  \beta (z e)^2 \kappa / (4\pi) 
$
gives 
\begin{eqnarray}
S_{\rm sub}[i\phi_{\rm cl}] &=& - \int_0^\infty d\xi \Bigg\{ 
{ \bar Z_p^2 g \over 2} \, \left( { du(\xi) \over d\xi} \right)^2 
\nonumber\\
&+& { \xi^2 \over g} \,  \left[ 
\exp\left\{ - { \bar Z_pg \over \xi} \, u(\xi) \right\} - 1 
+ { \bar Z_p g \over \xi} \, u(\xi) \, \right] \Bigg\} \,.
\nonumber\\
&&
\label{aaction}
\end{eqnarray}
Requiring that this new form of the action be stationary
produces the classical field equation 
\begin{equation}
- \bar Z_p g  \,  { d^2u(\xi) \over d\xi^2} =  \xi \,  \left[ 
\exp\left\{ - { \bar Z_pg \over \xi} \, u(\xi) \right\} - 1 
     \right] \,.
\label{equation}
\end{equation}
Note that the integral constraint (\ref{intc}) now reads
\begin{equation}
\int_0^\infty d\xi \, {\xi^2 \over g} \, \left[ 1 -
\exp\left\{ - { \bar Z_pg \over \xi} \, u(\xi) \right\} \right] = \bar Z_p \,.
\label{intcc}
\end{equation}

\subsection{Ion Sphere Model}

\begin{figure}  
\hspace{-2cm}
\includegraphics[width=7cm,height=4cm]{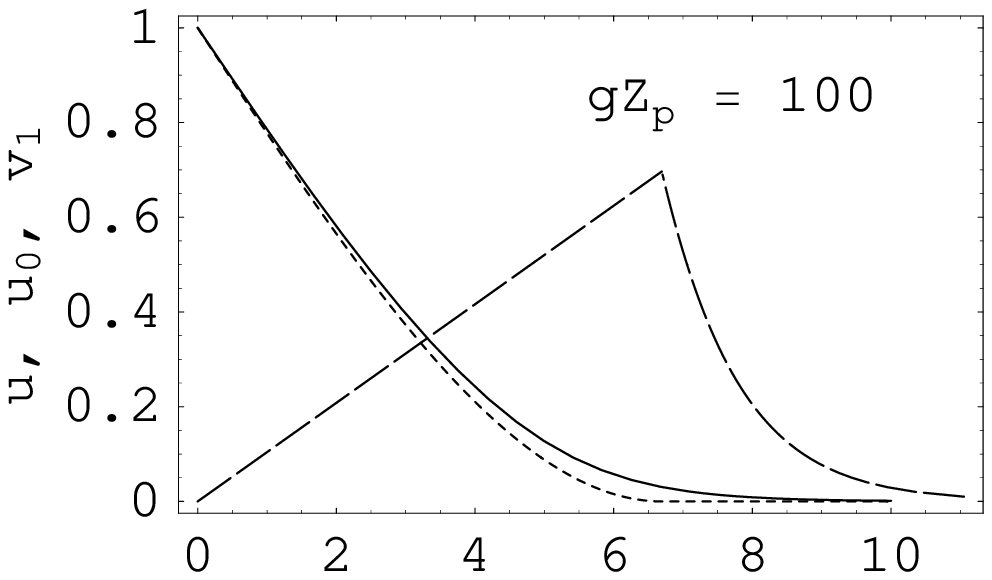}
\hspace{-2.3cm}  
{$\xi$}  
\caption{Numerical solution for $u(\xi)$ (solid line), 
ion sphere model $u_0(\xi)$ (short-dashed line), and the first 
correction $v_1$ (long-dashed line).  For $\xi > \xi_0$,
 $u_0 = 0$ ; here $\xi_0 = 6.694$.}
\label{uofxi}
\end{figure}

In the large $\bar Z_p$ limit which concerns us, the short distance
form (\ref{short}) (multiplied by $\beta ze $) is
large (compared to one) over a wide range of $|{\bf r}|$, 
and the Boltzmann factor
$ \exp\{ - \beta z e \phi_{\rm cl}({\bf r}) \} $ is quite small in this
range.  We are thus led to the ``ion sphere model'' brought forth some
time ago by Salpeter \cite{Sal}. This model makes the step-function 
approximation 
\begin{equation}
1 - \exp\left\{ - { \bar Z_pg \over \xi} \, u(\xi) \right\}
\simeq \theta\left( \xi_0 - \xi \right) \,.
\label{step}
\end{equation}
Placing this in the integral constraint (\ref{intcc}) determines
the ion sphere radius $\xi_0 = \kappa r_0 $ to be given by
\begin{equation}
 \xi_0^3 = 3 g \bar Z_p \,.
\end{equation}
In the ion sphere model,
the classical field equation (\ref{equation}) becomes
\begin{equation}
 \bar Z_p g  \,  { d^2 u_0(\xi) \over d\xi^2}  =
 \xi \,  \theta\left( \xi_0 - \xi \right) \,,
\label{sphereeq}
\end{equation}
and this has the solution, obeying the initial condition 
$u_0(0) =1$,
\begin{equation}
u_0(\xi) = \left\{
\begin{array}{ll}
   1 - \left( \xi / 2 \bar Z_pg \right) \left[ \xi_0^2 - {1\over3} \xi^2 
              \right] \,, & \mbox{$ \xi < \xi_0 \,, $} 
\\
       0  \,, & \mbox{$ \xi > \xi_0 \,. $}
\end{array}
\right.
\label{answer}
\end{equation}
Here the term linear in $\xi$, a solution of the homogeneous
equation, has been determined by the continuity at the ion sphere
surface, the condition that $u_0(\xi_0) =0$.  
[Without this constrain an additional
  $\delta(\xi-\xi_0)$ would appear on  the right-hand side of
  Eq.~(\ref{sphereeq}).] 
The nature of this ``ion-sphere'' solution $u_0(\xi)$ together with
the exact solution $u(\xi)$ obtained by the numerical integration of
Eq.~(\ref{equation}), as well as the first correction described below, 
are displayed in Fig.~\ref{uofxi}.  

We have appended the subscript $0$ to indicate that this is the
solution for the ion sphere model.  Placing this solution in the
new version (\ref{aaction}) of the action gives
\begin{equation}
- S_0\left[ i \phi_{\rm cl} \right] 
= {3 \bar Z_p \over10} \,  (3 g \bar Z_p)^{2/3}  - \bar Z_p \,.
\label{sphact}
\end{equation}
The final $-\bar Z_p$ that appears here comes from the 
$\left[\exp\left\{ - { \bar Z_pg \over \xi} \, u(\xi) \right\} 
   - 1 \, \right]$
term in the action (\ref{aaction}) along with the integral
constraint (\ref{intcc}).  This additional $-\bar Z_p$ simply adds a
constant to the chemical potential. Since a constant has no
dependence on the thermodynamic parameters, this addition has no
effect on the equation of state, the internal energy density, or
any other measurable thermodynamic quantity.  Moreover, 
the contributions of
such constants clearly cancels in the ratio (\ref{xrated}) that
yields the background plasma correction to the nuclear reaction
rate.  

\subsection{Ion Sphere Model Corrected}

To find the leading correction to the ion sphere model result, we
first cast the exact equations in a different form.  We start by
writing the full solution $u(\xi)$ as
\begin{equation}
u(\xi) = u_0(\xi) + { \xi_0 \over \bar Z_pg } \, v(\xi) \,,
\label{vdef}
\end{equation}
where $u_0(\xi)$ is the solution (\ref{answer}) to the ion sphere
model (\ref{sphereeq}).  The exact differential equation
(\ref{equation}) now reads
\begin{eqnarray}
&& - { d^2 v(\xi) \over d\xi^2 } = 
\nonumber\\
&&
{\xi \over \xi_0} \,
\left[ e^{- \bar Z_pg \, u_0(\xi) / \xi } \, \exp\left\{ 
  - \, { \xi_0 \over \xi} \, v(\xi) \right\} 
      - \theta\left( \xi - \xi_0 \right) \right] \,.
\nonumber\\
&&
\label{neweq}
\end{eqnarray}
Since $u_0(0) = 1$ is fixed (reflecting the presence of the
large, `impurity' point charge $Z_pe$), and since the solution 
must vanish at infinity, the proper solution to the non-linear 
differential equation (\ref{neweq}) is defined by the boundary  
conditions
\begin{equation}
v(0) = 0 \,, \qquad\qquad \xi \to \infty \,:  \qquad v(\xi) \to 0 \,.
\label{bc}
\end{equation}
On substituting the decomposition (\ref{vdef}) into the action
(\ref{aaction}), the cross term may be integrated by parts with
no end-point contributions in virtue of the boundary conditions 
(\ref{bc}) on  $v(\xi)$. We take advantage of this to move the
derivative of $v(\xi)$ over to act upon $u_0(\xi)$ so that we now have
$ d^2 u_0(\xi) / d\xi^2 $. Using Eq.~(\ref{sphereeq}) for this second
derivative and identifying the ion sphere part then gives
\begin{eqnarray}
S_{\rm sub}[i\phi_{\rm cl}] &=& S_0[i\phi_{\rm cl}] - 
{ \xi_0 \over g } \, \int_{\xi_0}^\infty d\xi \, \xi \, v(\xi)
\nonumber\\
&& \qquad
    - { \xi_0^2 \over 2g} \, \int_0^\infty d\xi \,
        \left( { dv(\xi) \over d\xi } \right)^2 \,.
\label{aaaction}
\end{eqnarray}

Thus far we have made no approximations.  To obtain the leading
correction to the ion sphere result, we note, as
 we have remarked before, that the factor
$ \exp\left\{ - { \bar Z_pg \over \xi} \, u_0(\xi) \right\} $
is very small for $ \xi < \xi_0 $, and so it may be evaluated by
expanding $u_0(\xi)$ about $\xi = \xi_0$. Using the result
(\ref{answer}), we find that the leading terms yield
\begin{equation} 
\exp\left\{ - { \bar Z_pg \over \xi} \, u_0(\xi) \right\} \simeq
\exp\left\{ - {1\over2} \, \left( \xi_0 - \xi \right)^2 \,
    \theta\left( \xi_0 - \xi \right)  \right\} \,.
\end{equation}
This approximation is valid for all $\xi$ because when $\xi$ is
somewhat smaller than $\xi_0$ and our expansion near the end
point breaks down, the argument in the exponent is so large that
the exponential function essentially vanishes. Indeed, since we
consider the limit in which $\xi_0$ is taken to be very large and
the Gaussian contribution is very narrow on the scale set by
$\xi_0$, we may approximate    
\begin{equation}
\exp\left\{ - { \bar Z_pg \over \xi} \, u_0(\xi) \right\} \simeq
\sqrt{ \pi \over 2} \, \delta\left( \xi - \xi_0 \right) +
   \theta\left(\xi - \xi_0 \right) \,.
\label{leader}
\end{equation}
Here the delta function accounts for the little piece of area
that the Gaussian provides near the ion sphere radius since
\begin{equation}
\int_0^\infty dx \, e^{- x^2 / 2} = \sqrt{ \pi \over 2} \,.
\end{equation}
With this approximation, an approximation that gives the leading
correction for the large $\bar Z_p g$ limit in which we work, 
Eq.~(\ref{neweq}) becomes
\begin{eqnarray}
- { d^2 v_1(\xi) \over d\xi^2} &=&    
   \sqrt{ \pi \over 2} \,  e^{-v_1(\xi_0)} \, 
     \,   \delta\left( \xi -\xi_0 \right)   
\nonumber\\
&& 
  +  \theta\left( \xi - \xi_0 \right) \, {\xi \over \xi_0 } \, 
   \left[ \exp\left\{ - { \xi_0 \over \xi } \, v_1(\xi) \right\}
                - 1 \right] \,.
\nonumber\\
&&
\label{diffeq}
\end{eqnarray}

It is easy to see that the first correction $v_1(\xi)$ does not
alter the integral constraint (\ref{intcc}).  Placing the
decomposition (\ref{vdef}) in the constraint (\ref{intcc}) and
using the leading-order form (\ref{leader}) together with
$v(\xi)$ replaced by $v_1(\xi)$ can be used to express the
putative change in the constraint (\ref{intcc}) in the form
\begin{eqnarray}
    \Delta \bar Z_p &=& - {\xi_0 \over g} \,
 \int_0^\infty d\xi \, \xi \,
 \Bigg\{ \sqrt{\pi \over 2} \, e^{-v_1(\xi_0)} \, 
                 \delta\left( \xi - \xi_0 \right) 
\nonumber\\
&+& 
  \theta\left( \xi - \xi_0 \right) \, { \xi  \over \xi_0} \,
\left[
    \exp\left\{- {\xi_0 \over \xi} v_1(\xi)\right\} 
- 1 \right] \Bigg\} \,.
\nonumber\\
&&
\end{eqnarray}
But Eq.~(\ref{diffeq}) and partial integration together with the
boundary conditions (\ref{bc}) now show that 
\begin{equation}
    \Delta \bar Z_p = {\xi_0 \over g} \, \int_0^\infty d\xi \,
          \xi \,  {d^2 v_1(\xi) \over d\xi^2 } = 0 \,.
\end{equation}

The $\delta(\xi - \xi_0)$ in Eq.~(\ref{diffeq}) requires that
\begin{equation}
\left. {d v_1(\xi) \over d\xi} \right|_{\xi = \xi_0 + 0} -
\left. {d v_1(\xi) \over d\xi} \right|_{\xi = \xi_0 - 0} 
=  - \sqrt{\pi \over 2} \, e^{ - \, v_1(\xi_0) } \,,
\label{discon}
\end{equation}
and
\begin{equation}
v_1(\xi_0 + 0) - v_1(\xi_0 -0) = 0 \,.
\label{con}
\end{equation}

Since
\begin{equation}
\xi < \xi_0 \,: \qquad\qquad  { d^2 v_1(\xi) \over d \xi^2} = 0 \,,
\end{equation}
and since $v(0) = 0$, we have
\begin{equation}
\xi < \xi_0 \,: \qquad\qquad  v_1(\xi) = c_1 \, \xi \,,
\end{equation}
where $c_1$ is a constant that is yet to be determined.  For large 
$\xi$, $v_1(\xi)$ is small and thus obeys the linearized version of
Eq.~(\ref{diffeq}), 
\begin{equation}
\xi \gg \xi_0 \,: \qquad\qquad 
        { d^2 v_1(\xi) \over d \xi^2} = v_1(\xi) \,,
\end{equation}
giving
\begin{equation}
\xi \gg \xi_0 \,: \qquad\qquad\quad
v_1(\xi) \sim e^{-\xi} \,.
\end{equation}
Since this damps rapidly on the scale set by
$  \xi_0 = (3 \bar Z_p g)^{1/3} \gg 1 $, the leading correction $v_1(\xi)$
that we seek is given by the solution to
\begin{equation}
\xi > \xi_0 \,: \qquad\qquad 
{ d^2 v_1(\xi) \over d \xi^2} = 1 - e^{- v_1(\xi) } \,,
\label{ddiffeq}
\end{equation}
which is 
the previous differential equation (\ref{diffeq}) in this region,
but with the explicit factors of $\xi/\xi_0$ and $\xi_0 / \xi$ 
replaced by $1$.  This new approximate second-order, non-linear
differential equation is akin  to a one-dimensional equation of motion
of a particle in a potential with $\xi$ playing the role of time, and 
$v_1(\xi)$ playing the role of position.  Thus there is an ``energy
constant  of the motion''.  Namely, if we multiply Eq.~(\ref{ddiffeq})
by $dv_1 / d\xi$, we obtain a total derivative with respect to $\xi$
whose integral gives 
\begin{equation}
{1 \over  2} \,  \left( {  d v_1(\xi) \over d\xi } \right)^2
- v_1(\xi) - e^{ - \, v_1(\xi) } = - 1 \,,
\end{equation}
where the constant $-1$ that appears on the right-hand side 
follows from  the limiting form as $ \xi \to \infty$.  It is easy to
show that 
\begin{equation}
e^{-v} + v -1 \ge 0 \,.
\end{equation}
Since  asymptotically $v_1(\xi)$ decreases when $\xi$ increases, 
we must choose the root 
\begin{equation}
 {  d v_1(\xi) \over d\xi } = - \sqrt{ 2 \, \left[ 
 e^{ - \, v_1(\xi) } +v_1(\xi) - 1 \right] } \,.
\label{slope}
\end{equation}

The different functional forms for $v_1(\xi)$ in the two regions 
$\xi < \xi_0$ and $\xi > \xi_0$ are joined by the continuity
constraint (\ref{con}), which we write simply as 
\begin{equation}
c_1 \, \xi_0 = v_1(\xi_0) \,,
\end{equation}
together with the slope jump (\ref{discon}) which, using
Eq.~(\ref{slope}), now requires that 
\begin{equation}
 \sqrt{ 2 \, \left[  e^{ - \, v_1(\xi_0) } +v_1(\xi_0) - 1 \right] } 
= \sqrt{\pi \over 2} \, e^{- v_1(\xi_0) } - {v_1(\xi_0) \over \xi_0} 
     \,.
\end{equation}
Since we require that $\xi_0 \gg 1$, the second term on the right-hand
side of this constraint may be neglected, which results in a
transcendental equation defining $v_1(\xi_0)$, whose solution is
\begin{equation}
v_1(\xi_0) = 0.6967 \cdots  \,.
\end{equation}

We are now in a position to evaluate the leading contribution to the
action (\ref{aaaction}).  Since $v_1(\xi)$ damps rapidly on the scale
set by $\xi_0$, in computing the leading term we can set $\xi= \xi_0$ 
in the integral that is linear  in $v_1(\xi)$. The
leading correction  is given by
\begin{equation}
S_{\rm reg}[i\phi_{\rm cl}] \simeq S_0[i\phi_{\rm cl}] + S_1 \,,
\end{equation}
in which
\begin{equation}
S_1 = - {\xi_0^2 \over g} \, {\cal C} \,,
\end{equation}
where
\begin{equation}
{\cal C} = \int_{\xi_0}^\infty d\xi \left\{ v_1(\xi) + {1\over2} \,
  \left( { d v_1(\xi) \over d\xi } \right)^2 \right\} \,.
\end{equation}
Here we have omitted the portion
\begin{eqnarray}
 \int_0^{\xi_0} d\xi \, {1\over2} \, 
\left( { d v_1(\xi) \over d\xi } \right)^2 &=&  
\int_0^{\xi_0} d\xi \, {1\over2} \, 
\left( {v_1(\xi_0) \over \xi_0 } \right)^2 
\nonumber\\
 &=&
{1\over2} \, {  v_1^2(\xi_0) \over \xi_0 } 
\end{eqnarray}
because it is parametrically smaller ---  it is of relative order
$1 / \xi_0$ to the leading terms that we retain.  We change variables
from $\xi$ to $v_1$ via
\begin{equation}
d\xi = \left( {dv_1 \over d\xi} \right)^{-1} \, dv_1 \,,
\end{equation}
and use the result (\ref{slope}) for the derivative.  Hence 
\begin{eqnarray}
{\cal C} &=& \int_0^{v_1(\xi_0)} { v_1 \, dv_1 \over 
 \sqrt{ 2 \, \left[ e^{ - \, v_1 } +v_1 - 1 \right] } }
\nonumber\\
&&
+ {1\over2} \,  \int_0^{v_1(\xi_0)} dv_1 \,
 \sqrt{ 2 \, \left[ e^{ - \, v_1 } +v_1 - 1 \right] }
\end{eqnarray}
is a pure number,
\begin{equation}
{\cal C} = 0.8499 \cdots \,.
\end{equation}

\begin{figure}
\hspace{-2cm}
\includegraphics[width=7cm,height=4cm]{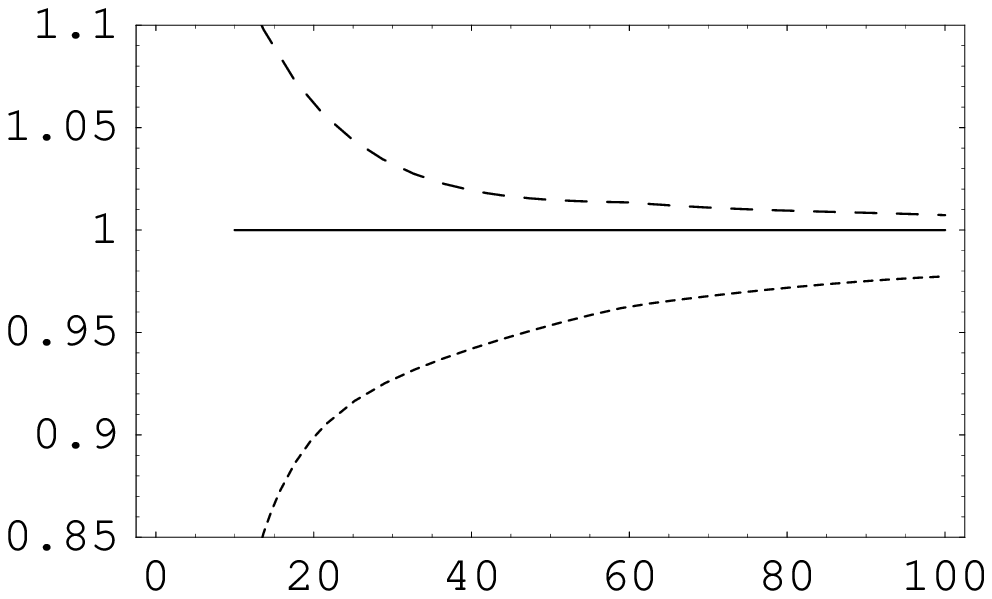}
\\ 
\vspace{-.5cm}
\hspace{1cm}
 {$g \bar Z_p$} 
\caption{Ratios of $S[i\phi_{\rm cl}] - \bar Z_p$ for 
the ion sphere model result (\ref{sphact}) [short-dashed line]
and the corrected ion sphere model (\ref{next}) [long-dashed line] 
to corresponding difference with the action (\ref{aaction}) for
the exact numerical solution $u(\xi)$ as functions of $g \bar Z_p$.}
\label{ratio}
\end{figure}

In summary, recalling that $\xi_0 = ( 3g\bar Z_p)^{1/3}$, we now find that
\begin{eqnarray}
 - [S_0 + S_1] + \bar Z_p 
&=& { 3 \bar Z_p \over 10 } \, \left( 3g\bar Z_p \right)^{2/3} \,
    \left\{ 1 + { 10 \, {\cal C} \over 3 g \bar Z_p }
   \right\} \,, 
\nonumber\\
&&
\label{next}
\end{eqnarray} 
with the leading correction to the ion sphere model exhibited as being
of relative order $ 1 / ( g\bar Z_p) $. Fig.~\ref{ratio} displays the exact 
numerical evaluation of the action compared with the ion sphere
approximation [the leading term in Eq.~(\ref{next})] and the corrected 
ion sphere model [the entire Eq.~(\ref{next})].

\subsection{One Loop}

The one-loop correction for the background plasma with no ``impurity''
ions present is given by \cite{foot}
\begin{equation}
{\rm Det}^{-1/2} \left[ - \nabla^2 + \kappa^2 \right] = 
\exp\left\{ \int (d^3{\bf r}) \, { \kappa^3 \over 12 \pi} \right\} \,.
\end{equation}
Since we assume that the charge $\bar Z_p$ of the ``impurity'' ions is so
large that not only $\bar Z_p \gg 1$, but also $ \bar Z_p g \gg
1$ as well, $ \kappa r_0 \gg 1$, and the 
ion sphere radius $r_0$ is large in comparison to the characteristic
distance scale for  spatial variation in the background plasma, the
Debye length $ \kappa^{-1}$.  In  this case, the term
\begin{equation}
\kappa^2 \, \exp\left\{ - \beta z e \phi({\bf r}) \right\} 
\end{equation}
in the one-loop determinant that enters into the background plasma
correction to the ``impurity'' number,
\begin{equation}
{\rm Det}^{-1/2} \left[ - \nabla^2 + \kappa^2 \,
      e^{- \beta ze \phi_{\rm cl} } \right]
\end{equation}
can be treated as being very slowly varying --- essentially a constant
--- except when it appears in a final volume integral. We conclude
that in this case of very strong coupling,
\begin{eqnarray}
&&{ {\rm Det}^{1/2} \left[ - \nabla^2 + \kappa^2 \right] \over 
{\rm Det}^{1/2} \left[ - \nabla^2 + \kappa^2 \,
      e^{- \beta ze \phi_{\rm cl} } \right]  }
\nonumber\\
&=&
\exp\left\{-  { \kappa^3 \over 12 \pi} \, \int (d^3{\bf r}) \,
\left[ 1 - \exp\left\{ - {3\over2} \beta ze \phi({\bf r})
  \right\} \right]
\right\} 
\nonumber\\
&=&
\exp\left\{-  { \kappa^3 \over 12 \pi} \, {4\pi \over 3} \, 
r_0^3 \right\} 
 = \exp\left\{ - {1\over3} \, g \bar Z_p \right\} \,,
\label{oneloop}
\end{eqnarray}
where in  the  second equality we have used the ion sphere  model that
gives the leading term for large $\bar Z_p$. 

This result is physically obvious.  The impurity ion of very high
$\bar Z_p$ carves out a hole of radius $r_0$ in the original,
background plasma, a hole that is a vacuum as far as the original
ions are concerned.  The original, background plasma is unchanged
outside this hole. This ion sphere picture gives the leading
terms for very large impurity charge $\bar Z_p$.  The corrections
that smooth out the sharp boundaries in this picture only produce
higher-order terms.  The original, background plasma had a
vanishing electrostatic potential everywhere, and the potential
in the ion sphere picture now vanishes outside the sphere of
radius $r_0$.  Thus the grand potential of the background plasma
is now reduced by the amount that was originally contained within
the sphere of radius $r_0$, and this is exactly what is stated to
one-loop order in Eq.(\ref{oneloop}).

This argument carries on to the higher loop terms as well, but we
shall now also sketch the application of the previous formal
manipulations to them as well.

\subsection{Higher Loops}

As shown in  detail in the paper of Brown and Yaffe \cite{BY},
$n$-loop terms in the expansion of the background plasma partition
function with no impurities present involve a factor of  
$ \kappa^2 \, \kappa^n $ which combines with other charge and
temperature factors to give dimensionless terms of the form
\begin{equation}
g^{n-1} \, \int (d^3{\bf r}) \, \kappa^3 \,.
\end{equation}
With the very high $\bar Z_p$ impurity ions present, each factor of
$\kappa$ is accompanied by 
$\exp\{ - (1/2) \beta e \, \phi_{\rm cl}({\bf r}) \} $
whose spatial variation can be neglected except in the final,
overall volume integral.  Thus, in the strong coupling limit of
the type that we have set, we have the order estimate
\begin{eqnarray}
&& n-{\rm loop} \,\,: 
\nonumber\\
&& \qquad g^{n-1} \kappa^3 \, \int (d^3{\bf r}) \,
\left[ 1 - \exp\left\{ - {n + 2 \over 2} \, \beta z e 
          \phi_{\rm cl}({\bf r}) \right\} \right]
\nonumber\\
&& \qquad\qquad 
\sim g^{n-1} \, \kappa^3 r_0^3 \sim g^n \, \bar Z_p \,.
\label{correct}
\end{eqnarray}

Again, since we assume that $g$ is sufficiently small so that
although $ g \bar Z_p \gg 1 $, $g^2 \, \bar Z_p \ll 1 $, all the
higher loop terms may be neglected.

In this discussion, we have glossed over the powers of $\ln g$
that enter into the higher-order terms as well as the quantum
corrections that can occur in higher orders. They vanish in our
strong coupling limit.

\acknowledgments

We thank Hugh E. DeWitt and Lawrence G. Yaffe for providing 
constructive comments on preliminary versions of this work.

\appendix*

\section{Rate Related to Previous Work}

We write the result (\ref{xrated}) in the form used by Brown,
Dooling, and Preston \cite{BDP} (BDP) which is not the notation
of DeWitt, Graboske, and Cooper \cite{DGC} (DGC).  In the grand
canonical methods employed by BDP, the temperature and chemical
potentials are the basic, fundamental parameters.  Thus, in this
grand canonical description, the effect of the background plasma
on nuclear reaction rates appears in terms of number changes with
the chemical potentials held fixed.  On the other hand, in the
canonical ensemble description employed by DGC, the temperature
and particle numbers are the basic, fundamental parameters.

To connect the two approaches, for the relevant case in which
``impurity'' ions $p$ are dilutely mixed in a background plasma,
we first note the general structure in the grand canonical
method.  Since the impurities are very dilute, the effect of the
background plasma on their number is entirely contained in the
first term of the fugacity expansion, the linear term in $ z_p =
\exp\{\beta\mu_p\}$.  In the free-particle limit where there is
no coupling of the impurities to the background plasma, the
impurity number density -- chemical potential connection reads
\begin{equation}
n^{(0)}_p = g_{s_p} \, \lambda_p^{-3} \, e^{\beta\mu_p} \,,
\end{equation}
where $g_{s_p}$ and $\lambda_p$ are the impurities' spin weight and 
thermal wavelength, respectively.  Thus the effect of the background
plasma appears as
\begin{equation}
n_p = n_p^{(0)} \,  e^{\Delta_p} =  
 g_{s_p} \, \lambda_p^{-3} \, e^{\beta\mu_p} \, e^{\Delta_p} \,,
\end{equation}
where we have chosen to write the plasma correction in terms of an
exponential.  The only feature  of the correction $\Delta_p$ that we
need note is that it is independent of the impurity fugacity 
$z_p$ since we are working in the $z_p \to 0$ limit.  In summary, the
correction in the grand canonical description appears as
\begin{equation}
 { n_p \over n_p^{(0)} } \, = e^{\Delta_p} \,,
\end{equation}
with the total number $N_P = n_p {\cal V}$, where ${\cal V}$ is
the volume of the system.

The grand canonical  partition function ${\cal Z}$ for the
complete system including the various impurity ions defines the
thermodynamic potential $\Omega(\beta,\mu)$ via
\begin{equation}
{\cal Z} = e^{-\beta \Omega} \,,
\end{equation}
and the particle number $N_a$ of species $a$ with chemical potential
$\mu_a$ is  given by
\begin{equation}
N_a = - { \partial \Omega \over \partial \mu_a } \,.
\end{equation}
Hence, since generically 
$\partial N^{(0)} / \partial \mu = \beta N^{(0)} $,
this can be integrated to produce
\begin{equation}
\Omega = \Omega_{\cal B} - {1\over\beta} \, {\sum}_p N_p^{(0)} \,
                              e^{\Delta_p} \,,
\label{OB}
\end{equation}
where $\Omega_{\cal B} $ is the thermodynamic potential of the
background plasma in the absence of the extra impurity ions and
where, as we have just shown, 
\begin{equation}
 N_p^{(0)} = N_p \, \exp\{ - \Delta_p \} \,. 
\end{equation}

The canonical partition function ${\cal Z}_N$ defines the
Helmholtz free energy $F(\beta, N)$ via
\begin{equation}
{\cal Z}_N = e^{-\beta F} \,,
\end{equation}
with the connection
\begin{equation}
F = \Omega + {\sum}_a \mu_a N_a \,.
\label{F}
\end{equation}
Since
\begin{equation}
\beta \mu_p = \ln \left(n_p^{(0)} \lambda_p^3 \, g_{s_p}^{-1}
 \right) \,,
\end{equation}
the Helmholtz free energy for a free gas of impurities is thus given
by 
\begin{equation}
\beta F_p^{(0)}\left(\beta, N_p^{(0)}\right) = N_p^{(0)} \, 
\left[ \ln\left( n_p^{(0)} \lambda_p^3 \, g_{s_p}^{-1} \right) 
- 1 \right] \,.
\end{equation}
The additional ionic impurities change the background plasma free
energy from 
\begin{equation}
F_{\cal B} = \Omega_{\cal B} + \sum_{a \ne p} \mu_a N_a \,, 
\label{FB}
\end{equation}
where the sum runs over all the particles in the plasma except
for the impurity ions, to
\begin{equation}
F = F_{\cal  B} + {\sum}_p F_p^{(0)}(\beta, N_p) 
                + {\sum}_p \, \Delta F_p \,.
\end{equation}
Using Eq's.~(\ref{FB}), (\ref{F}), and (\ref{OB}) produces
\begin{equation}
\beta \Delta F_p =  \beta \mu_P N_P - N_p^{(0)} \, e^{\Delta_p}
                  - \beta F_P^{(0)}(\beta, N_p) \,,
\end{equation}
and, since $\mu_p$ is fixed in terms of the free gas number
densities $ n_p^{(0)} = n_p \, \exp\{ - \Delta_p \} $, we find that
\begin{equation}
\beta \, \Delta F_p = - N_p \, \Delta_p \,.
\end{equation}
Thus, in the canonical ensemble approach employed by DGC \cite{DGC},
the previous number ratio is expressed in terms  of a Helmholtz free
energy change, 
\begin{equation}
{ n_p \over n_p^{(0)} } = \exp \left\{ - \beta 
                          { \Delta F_p \over N_p } \right\} \,.
\end{equation}
These authors sometimes write this in terms of a 
`chemical potential'.  However, within the grand canonical
description that we always employ, a chemical potential is an
independent variable that is not changed as interactions are
altered, and so in the context that we use this nomenclature
is not suitable..


\begin{thebibliography}{99}
\bibitem{foot}
        See Eq.~(2.79) of Brown and Yaffe \cite{BY} and the 
        discussion leading to that result.
\end{thebibliography}
\end{document}